\documentclass[aps,prd,preprint,groupedaddress,nofootinbib,amsmath,tightenlines]{revtex4}

\newcommand{\ul}[1]{\hspace{-0.1ex}\underline{\hspace{0.1ex} #1 \hspace{-0.1em}}\hspace{0.1em}}
\newcommand{\tauindicesstringI}{underlining }
\newcommand{\tauindicesstringII}{underlined }
\newcommand{\tauindicesstringIII}{``underlined'' }

\newcommand{\bydef}{\equiv} 
\newcommand{\R}{{\cal R}} 
\newcommand{\sss}{\scriptscriptstyle}
\newcommand{\sst}{\scriptstyle}
\newcommand{\ts}{\textstyle}
\newcommand{\ds}{\displaystyle}
\newcommand{\be}{\begin{equation}}
\newcommand{\ee}{\end{equation}}
\newcommand{\bea}{\begin{eqnarray}}
\newcommand{\eea}{\end{eqnarray}}
\newcommand{\bse}{\begin{subequations}}
\newcommand{\ese}{\end{subequations}}
\newcommand{\ifhat}{} 

\newcommand{\ket}[1]{|#1\rangle} 
\newcommand{\bra}[1]{\langle#1|} 
\newcommand{\scal}[2]{\langle#1|#2\rangle} 
\newcommand{\sandwich}[3]{\langle#1|#2|#3\rangle} 

\newcommand{\pic}{\pi} 
\newcommand{\pp}{\pi} 
\newcommand{\kk}{\kappa} 
\newcommand{\hpp}[1]{{\ifhat \pp_{#1}}} 
\newcommand{\hkk}[1]{{\ifhat \kk^{#1}}} 
\newcommand{\opppp}[1]{{\widehat{\left(#1\right)}_{\pp\pp}}} 
\newcommand{\okkkk}[1]{{\widehat{\left(#1\right)}_{\kk\kk}}} 
\newcommand{\oppkk}[1]{{\widehat{\left(#1\right)}_{\pp\kk}}} 
\newcommand{\smopppp}[1]{{\widehat{(#1)}_{\pp\pp}}} 
\newcommand{\smokkkk}[1]{{\widehat{(#1)}_{\kk\kk}}}
\newcommand{\smoppkk}[1]{{\widehat{(#1)}_{\pp\kk}}}

\newcommand{\DSx}{Y} 
\newcommand{\DS}[1]{Y_{\ul #1}} 

\newcommand{\ars}{{f}}  
\newcommand{\evds}[1]{{{\mathtt{y}}_{\ul #1}}} 

\newcommand{\arsphi}{{\phi_{\sss \!\ars}}} 
\newcommand{\arspsi}[1]{{{\psi_{\sss \! \ars}}_{\;\!\! \sst #1}}} 
\newcommand{\arsE}[1]{{{E_{\sss \! \ars}}_{\;\!\! \sst #1}}} 
\newcommand{\arsB}[1]{{{B_{\sss \! \ars}}_{\;\!\! \sst #1}}} 
\newcommand{\arsEprime}[1]{{{E'_{\sss \! \ars}}_{\;\!\! \sst #1}}} 
\newcommand{\arsBprime}[1]{{{B'_{\sss \! \ars}}_{\;\!\! \sst #1}}} 

\usepackage{bm}
\renewcommand{\vec}{\bm}

\begin{document}


\title{Algebraic Approach to the Duality Symmetry}


\author{Igor Salom}
\email{isalom@phy.bg.ac.yu}
\affiliation{Institute of Physics, P.O. Box 57, 11001 Belgrade,
Serbia and Montenegro}

\date{\today}

\begin{abstract}
The duality symmetry of free electromagnetic field is analyzed
within an algebraic approach. To this end, the conformal $c(1,3)$
algebra generators are expressed as operators quadratic in some
abstract operators $\hkk{\alpha}$ and $\hpp{\beta}$ which satisfy
Heisenberg algebra relations. It is then shown that the duality
generator can also be expressed in this manner. Standard issues
regarding duality are considered in such a framework. It is shown
that duality generator also generates chiral transformations, and
the conflict between duality and manifest Lorentz symmetry is
analyzed from the viewpoint of symmetry group greater then
conformal, in which duality generator appears as a natural part of
an $su(2)$ subalgebra.
\end{abstract}


\maketitle

\section{\label{sec1}Introduction}

Interest in the issue of duality symmetry has significantly
increased in recent years, due to its possible application in
non-perturbative regimes of various theories, in particular due to
its relation with string theories \cite{Coupling duality,
Strings}. Naturally, this has increased interest also in the first
of the dualities observed -- that of free field electric-magnetic
duality. This symmetry is interesting in its own right, not only
because it deals with something so common as propagation of light,
but also because, in spite of simplicity of its form, some aspects
of electric-magnetic duality still puzzle the physicists. First of
all, the manifest Lorentz covariance seems to be somehow
incompatible with electric-magnetic duality: standard covariant
Lagrangian of the free electromagnetic field does not possess
duality symmetry, as it changes sign upon transformations $E
\rightarrow - B$, $B \rightarrow E$. Usual attempts to maintain
Lorentz invariance together with the duality symmetry resorted
either to construction of non-polynomial action \cite{Non
polynomial action} or required an infinite set of fields
\cite{Infinite fields}. Following the idea of Majorana to write a
Dirac-like motion equation for photon where no non-physical
(gauge) freedom degrees would appear \cite{Majorana}, authors
\cite{Abreu} proposed a formulation where duality invariance is
obtained at the cost of introduction of auxiliary conjugate $E$
and $B$ fields and loss of manifest Lorentz covariance. Even with
abandoning of the manifest Lorentz covariance of action, there
remains a problem with the generator of (continuous) duality
transformations: such generator is either non-local \cite{Deser}
or requires introduction of auxiliary potential \cite{Schwartz}.
Another curious property of the duality symmetry, revealed in
attempts to formulate fermion-like formulation for the
electromagnetic field, is its connection with the chiral symmetry
\cite{Abreu}.

Unlike the analyses of the electromagnetic duality mentioned above
that start from Lagrangian formalism, Hamiltonian formalism, or
directly from equations of motion, in this paper we attempt to
give a kind of unified coverage of these topics from another
perspective, namely from algebraic approach. We exploit the fact
that conformal algebra (and thus also its Poincare subalgebra) is
contained as a subalgebra in algebra of operators quadratic in
some abstract operators $\hkk{\alpha}$ and $\hpp{\beta}$ which
satisfy Heisenberg algebra relations. In this context, we show
that the helicity operator can also be constructed as a quadratic
operator in terms of $\hkk{\alpha}$ and $\hpp{\alpha}$. In other
words, the helicity operator here appears on the same footing as
the conformal operators. We show that this operator generates
chiral transformations in the case of helicity $\pm \frac 12$
states and electromagnetic duality transformations in the case of
helicity $\pm 1$ states. The duality symmetry and the problem of
its incompatibility with manifest Lorentz symmetry are then
considered from a viewpoint of a larger mathematical symmetry that
appears in this formulation.

The paper is organized in the following way: in section \ref{sec2}
we construct conformal algebra in four dimensions $C(1,3)$ in
terms of the Heisenberg algebra operators. In section \ref{sec3}
we construct single particle Hilbert space and analyze action of
the helicity operator in subspaces of helicity $\pm \frac 12$ and
$\pm 1$. In section \ref{sec4} the issue of duality is considered
from the viewpoint of the larger symmetry group which embeds both
conformal group and duality transformation. Also, a possible
physical interpretation of this larger symmetry is considered.
Finally, section \ref{sec5} summarizes the results.

Throughout the text, Latin indices $i, j, k, \dots$ will take
values 1, 2 and 3, Greek indices from the beginning of alphabet
$\alpha, \beta, \dots$ will take values from 1 to 4 and will in
general denote Dirac-like spinor indices, while Greek indices from
the middle of alphabet $\mu, \nu, \dots$ will take values from 0
to 3, denoting Lorentz four-vector indices.

\section{\label{sec2}Constructing Poincare Algebra from Heisenberg Operators}

Let operators $\hkk{\alpha}$ and $\hpp{\alpha}$ satisfy Heisenberg
algebra in four dimensions: $[\hkk{\alpha}, \hpp{\beta}] = i
\delta_\beta^\alpha, [\hkk{\alpha}, \hkk{\beta}] = [\hpp{\alpha},
\hpp{\beta}] = 0$.\footnote{In spite of this, we stress that these
operators do not represent coordinates and momenta. Furthermore,
they will turn out to transform like Dirac spinors.} There are
three types of quadratic combinations of these operators:
quadratic in $\hkk{\alpha}$, quadratic in $\hpp{\alpha}$ and
mixed. Hermitian operators of each of these
kinds can be written in matrix notation, respectively as: %
\bea \okkkk{A} &\bydef& A_{\alpha\beta} \hkk{\alpha}\hkk{\beta}, \nonumber \\%
\opppp{A} &\bydef& A^{\alpha\beta} \hpp{\alpha}\hpp{\beta},
\label{quadratic operators} \\
\oppkk{A} &\bydef& A^\alpha_{\ \beta}
{\textstyle\frac12}\{\hpp{\alpha},\hkk{\beta}\}, \nonumber
\eea %
where $A$ is an arbitrary four by four real matrix.\footnote{A hat
sign over a matrix will be used to emphasize the difference
between the operator obtained from a matrix in the sense of
definition (\ref{quadratic operators}) and the matrix itself.}
However, due to commutativity of operators $\hkk{\alpha}$ among
themselves, and $\hpp{\alpha}$ among themselves, matrices
appearing in definitions $\okkkk{A}$ and $\opppp{A}$ are implied
to be symmetric.

Such quadratic operators form an algebra with commutation
relations easily derived from the Heisenberg algebra relations: %
\bea
  [\oppkk{A}, \oppkk{B}] &=& i\oppkk{[A,B]}, \nonumber  \\ {}
  [\oppkk{A}, \opppp{B}] &=& i\opppp{A B + B A^T}, \nonumber \\ {}
  [\oppkk{A}, \okkkk{B}] &=& -i\okkkk{A^T\! B+B A}, \label{second order algebra}  \\ {}
  [\opppp{A}, \okkkk{B}] &=& -4i\oppkk{A B},   \nonumber  \\ {}
  [\okkkk{A}, \okkkk{B}] &=& [\opppp{A}, \opppp{B}] = 0. \nonumber
\eea %

To reveal the Poincare subalgebra in this structure, first we
choose a set of six real matrices $\sigma_i$ and $\tau_{\ul i}$,
$i, \ul i = 1,2,3$ (four dimensional analogs of Pauli matrices) satisfying %
\be [\sigma_i, \sigma_j] = 2\:\! \varepsilon_{ijk} \sigma_k,\quad %
[\tau_{\ul i}, \tau_{\ul j}] = 2 \:\! \varepsilon_{\ul i \ul j \ul k} \tau_{\ul k},\quad %
[\sigma_i, \tau_{\ul j}] = 0, %
\label{sigma tau commutators} \ee %
as a basis of antisymmetric four by four real
matrices\footnote{One possible realization of such matrices is,
for example: $\sigma_1 = -i\sigma_y \times \sigma_x$, $\sigma_2 =
-i I_2 \times \sigma_y$, $\sigma_3 = -i\sigma_y \times \sigma_z$,
$\tau_1 = i\sigma_x \times \sigma_y$, $\tau_2 = -i \sigma_z \times
\sigma_y$, $\tau_3 = -i\sigma_y \times I_2$, where $\sigma_x$,
$\sigma_y$ and $\sigma_z$ are standard two dimensional Pauli
matrices and $I_2$ is a two dimensional unit matrix.} (we distinct
tau indices from sigma indices by \tauindicesstringI the former).
However, unlike Pauli matrices, these matrices are anti-hermitian,
satisfying $\sigma_i^2 = \tau_{\ul i}^2 = -1$. As a basis for
symmetric matrices we choose nine matrices $\alpha_{\ul ij} \bydef
\tau_{\ul i} \sigma_j$ and unit matrix denoted as $\alpha_0$. In
order to establish, later on, connection with standard notation,
we state one
corresponding representation of Dirac gamma matrices: %
\be \gamma_0 = i\tau_{\ul 2},\quad \gamma_i = \gamma_0 \;\!
\alpha_{\ul 3i} = i\tau_{\ul 1}\sigma_i, \quad \gamma_5 = -i
\gamma_0 \gamma_1 \gamma_2 \gamma_3 = i \tau_{\ul 3}.
\label{alpha and gamma connection}\ee %

Now, set of 36 operators %
\bea \Big\{\smoppkk{\tau_{\ul i}}, \oppkk{\sigma_j},
\oppkk{\alpha_0}, \smoppkk{\alpha_{\ul ij}}, \nonumber \\*
 \opppp{\alpha_0}, \smopppp{\alpha_{\ul ij}}, \okkkk{\alpha_0},
\smokkkk{\alpha_{\ul
ij}}\Big\} \label{Jeleb-conformal basis} \eea %
can be chosen as basis of algebra of quadratic operators.

Among the operators from this set, let us discard all those with
\tauindicesstringII index having values $\ul 1$ and $\ul 2$. This
resembles an idea to introduce a ``preferred tau direction''
(here, for concreteness, this ``direction'' was taken to be along
the third ``axis'') and to drop out every entity which has tau
indices but is not along this preferred direction. What we are
left with is a subalgebra isomorphic with conformal algebra C(1,3)
plus one additional generator that commutes with the rest of the
subalgebra. Now we introduce new notation for the
remaining generators: %
\bea & & \ifhat M_{ij} = \varepsilon_{ijk} \ifhat J_k \bydef
\varepsilon_{ijk} \oppkk{\displaystyle \frac{\sigma_k}{2}},
\nonumber \\ %
& & \ifhat M_{i0} = -\ifhat M_{0i} = \ifhat N_i \bydef
\oppkk{\displaystyle \frac{\alpha_{\ul 3 i}}{2}}, \quad  %
\ifhat D \bydef \oppkk{\displaystyle \frac{\alpha_0}{2}},
  \nonumber \\ %
& & \ifhat P_i \bydef \opppp{\displaystyle \frac{\alpha_{\ul 3
i}}{2}}, \quad \ifhat P_0 \bydef \opppp{\displaystyle \frac{
\alpha_0 }{2}}, \nonumber \\ %
& & \ifhat K_i \bydef \okkkk{\displaystyle \frac{\alpha_{\ul 3
i}}{2}}, \quad \ifhat K_0 \bydef -\okkkk{\displaystyle
\frac{\alpha_0}{2}}. \label{conformal identification} \eea

The additional remaining operator is %
\be \ifhat \DS{3} \bydef \oppkk{\ds \frac{\tau_{\ul 3}}{2}},
\label{helicity generator}\ee%
which commutes with all of the conformal generators. The operator
$\ifhat \DS{3}$ is in fact the helicity operator, as can be seen
from the following mathematical identity: %
\be \vec{\ifhat P} \cdot \vec{\ifhat J} = \ifhat P^0 \ifhat
\DS{3}. \label{helicity equation}\ee%
This can be verified most easily by a direct calculation, using
some concrete realization of the $\sigma$ and $\tau$ matrices.
[Note that $\vec P \cdot \vec J = -(P_1 J_1 + P_2 J_2 + P_3
J_3)$.]

We will show that in the subspace of helicity $\pm \frac 12$
states this identity directly turns into Dirac equation, while in
the subspace of helicity $\pm 1$ states the same identity
explicitly turns into a pair of Maxwell's equations. Furthermore,
we will demonstrate that the operator $\ifhat \DS{3}$ in the
former case generates chiral transformations while in the latter
case it generates electromagnetic duality transformations.

\section{\label{sec3}Single Particle Hilbert Space}

To this end we start by considering single particle Hilbert space,
which is analogue of the Hilbert space of non-relativistic quantum
mechanic with operators of coordinates and momenta replaced by
four pairs of ($\hkk{\alpha}$, $\hpp{\alpha}$) operators. In this
space basis vectors $\ket{p, \theta, \varphi, \evds{3}}$ exist
that are eigenstates of both helicity operator and spatial
momentum: %
\begin{eqnarray*} & & \ifhat \DS{3} \ket{p, \theta, \varphi, \evds{3}} =
\evds{3} \, \ket{p, \theta, \varphi, \evds{3}}, \\ %
& & \ifhat P^1 \ket{p, \theta, \varphi, \evds{3}} = p\,
\sin \theta \cos \varphi \, \ket{p, \theta, \varphi, \evds{3}}, \\ %
& & \ifhat P^2 \ket{p, \theta, \varphi, \evds{3}} = p\, \sin
\theta \sin \varphi \, \ket{p, \theta, \varphi, \evds{3}}, \\ %
& & \ifhat P^3 \ket{p, \theta, \varphi, \evds{3}} = p\, \cos
\theta \, \ket{p, \theta, \varphi, \evds{3}},
\end{eqnarray*} %
where $p \in [0,\infty),\theta \in [0, \pic], \varphi \in
[0,2\pic)$ and $\evds{3} = 0, \pm \frac 12, \pm 1, \dots$
Spherical expressions for momenta eigenvalues are more appropriate
here, since detailed calculation reveals that components of wave
functions that have half-odd integer values of helicity must be
$2\pic$ antiperiodic in angle $\varphi$ when expressed in this
basis.

Algebraic identity that is derived from definition (\ref{conformal identification}): %
\be \eta^{\mu\nu}\ifhat P_\mu \ifhat P_\nu = (\ifhat P_0)^2 -
(\ifhat
P_1)^2 - (\ifhat P_2)^2 - (\ifhat P_3)^2 = 0. \label{P squared}\ee %
implies that all states in this Hilbert space must be massless
(not surprising due to existence of conformal symmetry), so the
value of energy $\ifhat P_0$ of these states is simply the
magnitude of momentum. Normalization is chosen to provide that
under Lorentz transformations the states transform as:
$\ifhat\Lambda \ket{\vec p, \evds{3} = 0} =
\sqrt{\frac{p'^0}{p^0}} \, \ket{\vec{\Lambda p}, \evds{3} = 0}$.

Next, we define \emph{scalar field vectors}, namely Hilbert space
vectors that correspond to states of a single scalar particle
created at a given point $x$: %
\be \ket{\phi(x)} \bydef \int\limits_{\R^3} \frac{d^3\!p\
}{(2\pic)^{\sst 3/2}} \frac{1}{\sqrt{2 p^0}}\, e^{i p_\mu x^\mu}
\ket{\vec p, \evds{3} = 0}.
\label{scalar field vector} \ee %
Such vectors have simple Lorentz transformation properties:
$\ifhat \Lambda \ket{\phi(x)} = \ket{\phi(\Lambda x)}$.\footnote{
Uniqueness of these vectors can be better understood if the vector
$\ket{\phi(0)}$ is expressed in basis $\ket{\pp_1, \pp_2, \pp_3,
\pp_4}$ of operator $\hpp{\alpha}$ eigenstates, where its wave
function is simply a constant ($\ket{\phi(0)} \sim \int d^4\!\pp\
\ket{\pp_1, \pp_2, \pp_3, \pp_4}$). Action of any $\hkk{\alpha}$
operator on such state vanishes, so it is obviously invariant
under action of any operator of form $A^\alpha_{\ \beta}
\hpp{\alpha}\hkk{\beta}$ including Lorentz generators.}

For arbitrary Hilbert state $\ket{\ars}$ we define its
\emph{scalar field representation} as $\arsphi(x) \bydef
\scal{\phi(x)}{\ars}$. Direct calculation shows that action of
conformal generators in this representation (defined for arbitrary
generator $G$ as $G \arsphi(x) \bydef \sandwich{\phi(x)}{\ifhat
G}{\ars}$) reduce to standard formulas for classical fields.
Direct calculation also provides correct value for dilatation
charge for scalar field, something that is in the standard
approach usually inserted ``by hand'' to make the theory
dilatationally invariant (correct value of dilatation charge is
automatically obtained also for other fields, for example spinor
and helicity $\pm 1$ fields). The scalar field representation
function of arbitrary state defined in such way satisfies
Klain-Gordon equation, due to equality (\ref{P squared}).

The helicity $\pm \frac 12$ states are obtained by applying the
$\hpp{\alpha}$ operators to the scalar states. As these operators
transform under the spinor representation of Lorentz group (as can
be verified by calculating commutator $[M_{\mu\nu}, \hpp{\alpha}]$), states %
\be \ket{\psi_\alpha(x)} \bydef \sqrt 2 \hpp{\alpha} \ket{\phi(x)}
\label{spinor field vector} \ee %
transform like spinors. More precisely, function \be
\arspsi{\alpha}(x) \bydef \scal{\psi_\alpha(x)}{\ars},
\label{spinor field representation} \ee %
that we are going to call \emph{spinor field representation} of a
state $\ket{\ars}$, transforms as a classical spinor field, under
both Lorentz and conformal group. In particular, we find: %
\bea P_\mu \arspsi{\alpha}(x) &=& \partial_\mu \arspsi{\beta}(x),
\label{P and M in spinor field representation}\\
M_{\mu\nu} \arspsi{\alpha}(x) &=& i \left( (x_\mu
\partial_\nu - x_\nu
\partial_\mu)\delta^\beta_\alpha + (\sigma_{\mu\nu})_\alpha^{\
\beta}\right) \arspsi{\beta}(x) \nonumber
\eea %
where, as usually, $\sigma_{\mu\nu} \bydef \frac 14 [\gamma_\mu,
\gamma_\nu]$.

We now find action of the $\ifhat \DS{3}$ operator in the spinor
field representations (in the scalar field case this
operator trivially reduces to zero): %
\bea \DS{3} \arspsi{\alpha}(x) &\bydef&
\sandwich{\psi_\alpha(x)}{\ifhat \DS{3}}{\ars} \nonumber \\
&=& i (\frac{\tau_{\ul 3}}{2})_\alpha^{\ \beta} \arspsi{\beta}(x)
= {\ts \frac{1}{2}}(\gamma_5)_\alpha^{\ \beta} \arspsi{\beta}(x).
\label{DS3 in spinor field representation}\eea %

So, in the spinor field representation operator $\ifhat \DS{3}$
effectively turns into the chiral charge matrix $\gamma_5$.

To demonstrate that function $\arspsi{\alpha}(x)$ behaves like
massless Dirac field we will show that the Dirac equation is
satisfied. Using results (\ref{P and M in spinor field
representation}) and (\ref{DS3 in spinor field representation})
mathematical identity (\ref{helicity equation}) directly leads to
massless Dirac equation for spinor field functions: %
\bea 0 \!\! &=& \! \bra{\psi_\alpha(x)} \ifhat P_0 \ifhat \DS{3}
+ \sum_i \ifhat P_i \ifhat J_i \ket{\ars} \nonumber \\ %
&=& \! \bigg({\ts \frac i2} (\gamma_5)_\alpha^{\ \beta} \partial_0
\nonumber + \sum_{ijk} i \partial_i \varepsilon_{ijk} \Big(i x_j
\partial_k\, \delta_\alpha^\beta + {\ts \frac
i2}(\sigma_{jk})_\alpha^{\ \beta}
\Big)\bigg) \arspsi{\beta} \nonumber \\ %
&=& \! \Big({\ts \frac i2} (\gamma_5)_\alpha^{\ \beta} \partial_0
- \sum_i {\ts \frac i2}(\gamma_5 \gamma_0 \gamma_i)_\alpha^{\
\beta}
\partial_i \Big) \arspsi{\beta}.
\eea%
Suppressing the spinorial indices and multiplying by $2 \gamma_0
\gamma_5$ from the left, we obtain the massless Dirac equation
in its standard form: %
\be i \gamma^\mu \partial_\mu \arspsi{}(x) = 0.
\label{massless Dirac equation}\ee %

Just as we applied $\hpp{\alpha}$ operators once to scalar field
vectors in order to obtain basis for helicity $\pm \frac 12$
states, we can apply these operators twice, i.e.\ $\hpp{\alpha}
\hpp{\beta} \ket{\phi(x)}$ to obtain basis for field
representation of helicity $\pm 1$ states. However, four out of
ten possible quadratic combinations of $\hpp{\alpha} \hpp{\beta}$
will not change helicity -- these are ones corresponding to
momenta (since momenta commute with $\ifhat \DS{3}$). Using the
six remaining combinations we define $E$ and $B$ vectors: %
\bse \label{E and B Hilbert vectors} %
\bea \ket{E_i(x)} &\bydef& \opppp{\alpha_{\ul 1i}} \ket{\phi(x)},
\\ \ket{B_i(x)} &\bydef& - \opppp{\alpha_{\ul 2i}} \ket{\phi(x)}.
\eea \ese %
Here linear combination $\ket{E_i(x)} \mp i \ket{B_i(x)}$ has
helicity value $\pm 1$.

Corresponding $E$ and $B$ field representation functions: %
\be \arsE{i}(x) \bydef \scal{E_i(x)}{\ars}, \qquad
\arsB{i}(x) \bydef \scal{B_i(x)}{\ars}, \label{E and B functions} \ee %
have the same Lorentz transformation properties as electric and
magnetic fields, respectively.\footnote{It is true that so defined
$\arsE{i}$ and $\arsB{i}$ functions can take complex values, which
is not a property of standard electric and magnetic fields.
However, this is hardly avoidable in one first quantization
approach like this, where $E$ and $B$ functions are understood to
play role of a photon wave function (the idea that physical $E$
and $B$ fields instead of potential $A_\mu$ should be related to
photon wave function is usually attributed to Majorana
\cite{Majorana}).}

For representation of $\ifhat \DS{3}$ operator we find: %
\bea \DS{3} \arsE{i}(x) \bydef \sandwich{E_i(x)}{\ifhat
\DS{3}}{\ars} = i \arsB{i}(x), \nonumber \\ \DS{3} \arsB{i}(x)
\bydef \sandwich{B_i(x)}{\ifhat \DS{3}}{\ars} = -i \arsE{i}(x)
\label{DS3 in E and B field representation}
\eea %
that results in following finite transformations: %
\bea \arsE{i}(x) & \longrightarrow& \arsEprime{i}(x)
= \arsE{i}(x) \cos \phi - \arsB{i}(x) \sin \phi, \nonumber \\
\arsB{i}(x) &\longrightarrow& \arsBprime{i}(x) = \arsE{i}(x) \sin
\phi +\arsB{i}(x) \cos \phi \label{DS3 in
Maxwell field representation} \eea %
corresponding to change of state given by $\ket{\ars}
\longrightarrow \exp(i \phi \ifhat \DS{3}) \ket{\ars}$.

In the similar manner as the Dirac equation was derived from the
helicity identity (\ref{helicity equation}) in the spinor field
representation, now a pair of Maxwell's equations is derived
from the same identity: %
\bea & \bra{E_i(x)} \ifhat P_j \ifhat J_j \ket{\ars} = {}
{}-\bra{E_i(x)} \ifhat P_0 \ifhat \DS{3} \ket{\ars} \nonumber \\ %
& \Rightarrow {(s_j)}_{ik}
\partial_j \arsE{k} = - \partial_0 \arsB{i} \Rightarrow
\varepsilon_{ijk} \partial_j \arsE{k} = -
\partial_0 \arsB{i}, \nonumber \\ %
& \bra{B_i(x)} \ifhat P_j \ifhat J_j \ket{\ars} = -\bra{B_i(x)}
\ifhat P_0 \ifhat \DS{3} \ket{\ars} \nonumber \\ %
& \Rightarrow {(s_j)}_{ik}
\partial_j \arsB{k} = \partial_0 \arsE{i} \Rightarrow
\varepsilon_{ijk} \partial_j \arsB{k} = \partial_0 \arsE{i}. %
\label{Maxwell equations} \eea %
Summation over repeated indices is implied and matrices $s_j$ are
matrices generating rotations in three dimensional vector
representation of rotation group. Matrix notation of intermediate
results in (\ref{Maxwell equations}) is the essence of what is
sometimes called fermion-like formulation for electromagnetic
field \cite{Majorana, Abreu}. (One can draw closer parallels to
Majorana original fermion-like formulation by expressing these
results in terms of linear combinations $\arsE{i}(x) \pm
i\arsB{i}(x)$ of definite helicity.)

The other two Maxwell equations can be derived from mathematical
identity\footnote{Fact that equations of motion are simply
mathematical tautologies is a nice characteristic of this approach.} %
\be \sum_i \opppp{\alpha_{\ul ji}} \opppp{\alpha_{\ul ki}} =
\delta_{\ul j\ul k} \opppp{\alpha_0}^2,
\label{PP orthogonality relations} \ee %
by taking consecutively $\ul j = 3, \ul k = 1$ and $\ul j = 3, \ul
k = 2$ (Choice $\ul j = 3, \ul k = 3$ gives identity $P^\mu P_\nu
= 0$).

Since the functions $\arsE{i}(x)$ and $\arsB{i}(x)$ satisfy free
Maxwell equations, we may recognize transformations (\ref{DS3 in
Maxwell field representation}) as (continuous) duality
transformations of free electromagnetic field.

These results can easily be generalized to field representations
of arbitrary helicity, and thus equivalent of duality (chirality)
can also be defined for these cases. The same is true for
generalization of the massless Dirac equation. It is clearly seen
here that this equation is essentially just a helicity eigenvalue
problem.

To conclude this section we note that in this approach duality
generator appears as a well defined operator, a part of the
starting algebra, whose action on free $E$ and $B$ fields
(\ref{DS3 in Maxwell field representation}) is localized in
space-time. As a matter in fact this operator is also the helicity
operator as well as the fermion chiral symmetry generator. The
result that helicity operator (in massless case) generates duality
transformations essentially agrees with findings of
\cite{Girotti}, in spite of the differences in approach. The
connection of duality and chirality is also not new. Conclusion
that ``duality is a kind of chirality'' can be found in ref.\
\cite{Abreu}. However, the authors used there the phrase ``kind
of'' since such a conclusion was based simply on the fact that
they implemented duality transformations using a matrix that
anticommutes with analogues of Dirac $\gamma_\mu$ matrices
appearing in a fermion-like formulation for the electromagnetic
field. The connection of duality with chiral symmetry is more
clearly established with two relations (\ref{DS3 in spinor field
representation}) and (\ref{DS3 in E and B field representation})
and in this framework it turns out that ``duality is chirality'',
i.e.\ both symmetries are generated by the same operator.

All of these conclusions were derived for the massless and free
field case. Nevertheless, let us briefly consider more general
case and, in the light of this connection of chirality with
duality, consider the following. It is well known that the mass
term of the Dirac equation %
\be i \gamma^\mu \partial_\mu \Psi(x) = m \Psi(x) \label{mass1} \ee %
breaks chiral invariance, since, for example, a ``$90^\circ$
chiral rotation'' would require the standard mass term to be
replaced with
mass term of the form $i\gamma_5 m$: %
\be i \gamma^\mu \partial_\mu \Psi(x) = i \gamma_5 m \Psi(x).
\label{mass2}  \ee %
On the other hand the same symmetry transformation (appearing now
as a duality transformation) is expected to turn an electric
charge into a magnetic charge. As a consequence of this somewhat
simplified consideration a possibility arises that some relation
might exist between magnetic charges and $i\gamma_5 m$ type of the
mass term.\footnote{A similar mass term appears, for example, in
papers of Raspini \cite{Raspini} and Dvoeglazov \cite{Dvoeglazov},
but without the imaginary constant. Lack of this constant in their
case introduces exotic massless and tachyonic solutions, which do
not occur otherwise.} As the discussion of fields with sources is
out of the scope of this paper, we shall not discuss such a
possibility in more details.

\section{\label{sec4}Symmetries Beyond Conformal}

The algebraic formulation of this work provides a different way to
understand duality and also offers a new perspective on conflict
between Lorentz covariance and duality symmetry. To fully
demonstrate this it is instructive to take a broader viewpoint.
Namely, let us recall that conformal algebra, when expressed using
Heisenberg operators (\ref{conformal identification}), is from a
mathematical point of view natural part of a greater algebra of
all quadratic operators in $\hpp{\alpha}$ and $\hkk{\alpha}$
(\ref{Jeleb-conformal basis}). If we include the rest of these
operators in consideration, it is first noticed that duality
generator $\ifhat \DS{3}$ mathematically belongs to an $su(2)$
subalgebra generated by $\ifhat \DS{1} \bydef \smoppkk{\ds
\frac{\tau_{\ul 1}}{2}}$, $\ifhat \DS{2} \bydef \smoppkk{\ds
\frac{\tau_{\ul 2}}{2}}$ and $\ifhat \DS{3}$. This is an unusual
result. This $su(2)$ algebra commuting with rotational generators
we designate as \emph{dual spin} algebra. The dual-spin generators
and rotation generators, together with nine operators $\ifhat
N_{\ul ij} \bydef \smoppkk{\frac{\alpha_{\ul i j}}{2}}$ [three of
these with $\ul i = 3$ are boosts according to (\ref{conformal
identification})] form another algebra [$sl(4,R)$ isomorphic],
which is an extension of the Lorentz algebra.

On the other hand four momenta $\ifhat P_\mu$ naturally fit into a
set of ten operators quadratic in $\hpp{\alpha}$: $\ifhat P_0$ and
nine operators $\ifhat P_{\ul ij} \bydef
\smopppp{\frac{\alpha_{\ul ij}}{2}}$. Just as four-momentum
transforms under irreducible vector representation of Lorentz
group these ten operators belong to an irreducible representation
of the extended Lorentz group. Under symmetry reduction from this
extended Lorentz group to Lorentz group, this ten dimensional
representation decomposes into irreducible representations of
Lorentz group as $(\frac 12, \frac 12) \oplus (1, 0) \oplus (0,
1)$. Here, components of four momenta $\ifhat P_\mu$ belong to
$(\frac 12, \frac 12)$ subspace, while six operators $P_{\ul 11}$,
$P_{\ul 12}$, $P_{\ul 13}$, $P_{\ul 21}$, $P_{\ul 22}$, $P_{\ul
23}$ transform under $(1, 0) \oplus (0, 1)$ Lorentz
representation, i.e.\ they transform as an antisymmetric second
rank Lorentz tensor. This makes sense when we repeat our
definition of $\arsE{i}(x)$ and $\arsB{i}(x)$ functions (\ref{E
and B Hilbert vectors}, \ref{E and B functions}), here written as: %
\be \arsE{i}(x) \bydef 2\sandwich{\arsphi(x)}{\ifhat P_{\ul
1i}}{\ars}, \quad \arsB{i}(x) \bydef
-2\sandwich{\arsphi(x)}{\ifhat P_{\ul 2i}}{\ars}. \label{E and B
functions 2} \ee %

From this aspect duality generator, being the third component of
the dual spin, rotates electric field that is ``along the first
dual axis'' into magnetic field that is ``along the second dual
axis''.\footnote{If the rotation around the ``third dual axis'' in
positive direction is introduced with a minus sign in exponential
[$\ket{\ars} \rightarrow \exp(-i \phi \ifhat \DSx_{3})
\ket{\ars}$] then, strictly speaking, the results (\ref{DS3 in
Maxwell field representation}) and (\ref{mass2}) correspond to
dual rotations in negative direction.} It is clear that $(\vec
E)^2 + (\vec B)^2$ is the only dually and rotationally invariant
quadratic function of $E$ and $B$, whereas product $2 \vec E \cdot
\vec B$ and the standard Lagrangian $(\vec E)^2 - (\vec B)^2$
transform into each other. This is completely analogous to the
case of, for example, spatial momenta where $p_x^2 + p_y^2$ is
invariant under rotations around $z$ axis, whereas under the same
rotations functions $p_x^2 - p_y^2$ and $2 p_x p_y$ transform into
each other. Existence of the two mass terms (\ref{mass1}) and
(\ref{mass2}) can be understood also on the basis of the extended
Lorentz group: these mass terms belong to a six dimensional
representation of the extended Lorentz group decomposing into
$(\frac 12, \frac 12) \oplus (0, 0) \oplus (0,0)$ Lorentz
representations where one of the two Lorentz scalars is ``along
the first dual axis'' and the other is ``along the second dual
axis''. Furthermore, we conclude that if tensor components
$F_{\mu\nu} \bydef \partial_\mu A_\nu -
\partial_\nu A_\mu$ are to transform in the same way as six
entities $(E, B)$ (\ref{E and B functions 2}) with respect not
only to Lorentz but also and to duality transformations, potential
$A_\mu$ needs additional \tauindicesstringIII dual spin indices.

Apart from the operators already mention in this section, the full
algebra of quadratic Heisenberg operators also includes an
extended set of ten ``conformal-like'' operators $\ifhat K_0$,
$\ifhat K_{\ul ij} \bydef \smokkkk{\frac{\alpha_{\ul ij}}{2}}$ and
one dilatation generator $\ifhat D$ already defined in
(\ref{conformal identification}). This whole, 36 dimensional
algebra can be mathematically seen as an extension of conformal
algebra and it is isomorphic with symplectic algebra in four
dimensions.

A natural question in this context is about the interpretation of
the new operators appearing in this extended conformal algebra. So
far, this bigger algebra was essentially considered only as a
mathematical extension appropriate for this formulation. Now we
briefly discuss a possible physical interpretation of the whole
algebra as a space-time symmetry.

If this is to be a physical symmetry, it can only be a broken one.
It is interesting that a relatively simple form of symmetry
breaking can reduce this big symmetry group (with 36 generators)
to conformal symmetry (15 generators) multiplied by $U(1)$ group
of duality transformations. It is sufficient that dual-spin
$SU(2)$ group is strongly broken to its $U(1)$ subgroup, for
example by some effective interaction with potential of the form
$const \cdot (\DS{3})^2$, where the multiplying constant is
sufficiently large (let us say of order of Planck mass divided by
Planck constant squared). Such a term would effectively reduce low
energy physics to the Hilbert space subspace characterized with
$\evds{3} = 0$ while the remaining space-time symmetry would be
conformal group. However, symmetry breaking provided by such a
simple term is inadequate because of other reasons (for example,
non scalar states in this simplest formulation would acquire
enormous masses). Nevertheless this possibility incorporates one,
to our opinion, appealing idea that fundamental postulated
symmetry of space-time fits into some clear mathematical pattern
(here it is algebra of all quadratic Heisenberg operators), rather
then to be just given by not so simple structural constants (as in
the case of Poincare symmetry). It might be argued that the former
case contains ``less information'' and thus that it is favored by
Occam's razor. Due to remote associations with the standard model,
it is also potentially interesting that the necessary symmetry
breaking is connected with an $SU(2)$ to $U(1)$ breaking, where
the group in question is related with chirality. We shall not
dwell any longer on this topic, remarking only that if any
realistic model is sought with the symmetry given by algebra of
operators (\ref{second order algebra}, \ref{Jeleb-conformal
basis}), it seems more appropriate to start with a trilinear
generalization of Heisenberg commutation relations of the form $[[
\hpp{\alpha}, \hpp{\beta}]_+, \hpp{\gamma}]_- = 0$, $[[
\hkk{\alpha}, \hkk{\beta}]_+, \hkk{\gamma}]_- = 0$, $[[
\hkk{\alpha}, \hpp{\beta}]_+, \hkk{\gamma}]_- =
2i\delta^\gamma_\beta \hkk{\alpha}$,  $[[ \hpp{\alpha},
\hkk{\beta}]_+, \hpp{\gamma}]_- = 2i\delta_\gamma^\beta
\hpp{\alpha}$, (a graded algebra isomorphic to four dimensional
para-Bose algebra whose Green's representations \cite{Green} seem
adequate for representing multiparticle states).

At the end, it should be mentioned that understanding the whole
algebra as the one generating physical symmetry would have
consequences also on the issue of duality. Invariance of
Lagrangian terms and matching of transformation properties [for
example of $\partial_\mu A_\nu -
\partial_\nu A_\mu$ and $E ,B$ given by (\ref{E and B functions
2})] should be then considered while regarding the full extended
conformal group.

\section{\label{sec5}Conclusion}

The conformal generators were expressed in this paper as quadratic
functions of operators satisfying Heisenberg algebra. In such a
formulation it turned out to be possible to express the helicity
operator in the same way, putting it on the same level with the
conformal algebra generators. This helicity operator upon action
on helicity $\pm \frac 12$ states behaved as chirality generator
and upon action on combination of helicity $\pm 1$ states behaved
as duality generator. Thus we put the parallels of duality and
chirality \cite{Abreu} on stronger grounds. As a convenient
feature of this approach we also demonstrated that massless motion
equations without sources appear here as mathematical tautologies.
Next we pointed out that an unusual perspective on concept of
duality is obtained from the viewpoint of complete algebra of
quadratic Heisenberg operators, where the duality generator
naturally fits into one $su(2)$ algebra. Finally, this large
algebra was briefly discussed as a candidate for a physical
symmetry of universe. Although the most simple symmetry breaking
mechanism mentioned in the section 4 does not meet some of the
basic experimental requirements, it was concluded that a simple
symmetry breaking of $SU(2)$ dual-spin group to its $U(1)$ duality
subgroup could reduce space-time symmetry to conformal symmetry.


\begin{acknowledgments}
The author of this paper is deeply indebted to Prof.\ Djordje
Sijacki for numerous suggestions and helpful comments. The work is
supported in part by Serbian Ministry of Science and Environmental
Protection, under contract No.\ 141036.
\end{acknowledgments}


\end{document}
%